\documentstyle[pre,aps,twocolumn,psfig]{revtex}     

\newcommand{\Lmat}[1]{{{\bf L}_{#1}}}	   
\newcommand{\Bmat}[1]{{{\bf B}_{#1}}}	   

\newcommand{\field}{\phi}

\newcommand{\rf}     [1] {~\cite{#1}}
\newcommand{\refref} [1] {ref.~\cite{#1}}
\newcommand{\refrefs}[1] {refs.~\cite{#1}}

\newcommand{\refeq}  [1] {(\ref{#1})}

\newcommand{\reffig} [1] {fig.~\ref{#1}}

\newcommand{\reftab} [1] {table~\ref{#1}}

\newcommand{\beq}{\begin{equation}}
\newcommand{\continue}{\nonumber \\ }
\newcommand{\nnu}{\nonumber}
\newcommand{\eeq}{\end{equation}}
\newcommand{\ee}[1] {\label{#1} \end{equation}}
\newcommand{\bea}{\begin{eqnarray}}
\newcommand{\ceq}{\nonumber \\ & & }
\newcommand{\eea}{\end{eqnarray}}
\newcommand{\barr}{\begin{array}}
\newcommand{\earr}{\end{array}}

\newcommand{\evOper}{evolution oper\-ator}

\newcommand{\FPoper}{Perron-Frobenius oper\-ator} 

\newcommand{\fd}{spec\-tral det\-er\-min\-ant}

\newcommand{\defeq}{=}		

\newcommand{\pde}{\partial}

\renewcommand{\det}{\mbox{\rm det}}

\newcommand{\tr}{{\rm tr}\, }

\newcommand{\Lop}{{\cal L}}	   

\newcommand{\ExpaEig}{\Lambda}	   
\newcommand{\eigenvL}{{\nu}}       
\newcommand{\eigCond}{F}           

\newcommand{\cl}[1]{{n_{#1}}}	

\begin{document}
\draft{
\title{Spectrum of stochastic evolution operators: local matrix representation
approach}

\author{Predrag Cvitanovi\'c and Niels S\o ndergaard}
\address{
Department of Physics \&\ Astronomy, Northwestern University \\
2145 Sheridan Road, Evanston, Illinois 60208}
\author{Gergely Palla and G\'abor Vattay}
\address{Department of Physics of Complex Systems, E\"otv\"os University\\
P\'azm\'any P\'eter s\'etany 1/A,
H-1117 Budapest, Hungary   } 
\author{C.P. Dettmann}
\address{Rockefeller University   \\
1230 York Ave.,
New York City, NY 10021
	}

\date{\today}

\maketitle

\begin{abstract}
A matrix representation of the \evOper\  associated with
a nonlinear stochastic flow with additive noise
is used to compute its spectrum.
In the weak noise limit a perturbative expansion for the spectrum
is formulated in terms of local matrix representations of the \evOper\ centered 
on classical periodic orbits. 
The evaluation of perturbative corrections is easier to implement
in this framework than in the standard Feynman diagram perturbation theory.
The result are perturbative 
corrections to a stochastic analog of the Gutzwiller semiclassical
{\fd} computed to several orders beyond
what has so far been attainable in stochastic and quantum-mechanical
applications.
\end{abstract}
\pacs{02.50.Ey, 03.20.+i, 03.65.Sq, 05.40.+j, 05.45.+b}
}

\section{Introduction}

Any dynamical evolution that occurs in nature is affected by noise.
In a neuronal system the noise might be comparable in magnitude to
purported underlying deterministic dynamics; in celestial 
mechanics the degrees of freedom omitted from a particular
set of equations may be accounted for by very weak noise.
Our task here and in two preceding papers\rf{noisy_Fred,conjug_Fred}
is to systematically account for the effects of noise on measurable
properties such as dynamical averages\cite{bene} in classical chaotic dynamical
systems.

The theory is also closely related to the semiclassical expansions based
on Gutzwiller's formula for the trace in terms of classical periodic
orbits\rf{gutbook} in that both are perturbative theories (in the noise
strength or $\hbar$) derived from saddlepoint expansions of a path
integral containing a Cantor set of unstable stationary points (typically
periodic orbits).  The analogy with quantum mechanics and
field theory has been made explicit in\rf{noisy_Fred} where
Feynman diagrams were used to find the lowest nontrivial noise corrections.
Unfortunately like its quantum counterpart, the Feynman diagram method
for stochastic dynamics quickly becomes unwieldy at higher orders; rather
than applying it directly we turn the argument around and suggest that
the more efficient recent approaches of\rf{conjug_Fred} and the present
paper be applied to difficult perturbative problems of quantum mechanics
and field theory.

An elegant method, inspired by the classical perturbation theory of
celestial mechanics, is that of smooth conjugations\rf{conjug_Fred}.
In this approach
the neighborhood of each saddlepoint is flattened by an appropriate
coordinate transformation, so the focus shifts from the original dynamics
to the properties of the transformations involved.  An elementary example
is the Ulam map $f(x)=4x(1-x)$ which is solved exactly by the transformation
$x=\sin^2(\pi\theta/2)$ leading to the piecewise linear tent map
$f(\theta)=1-|1-2\theta|$.  In general there is no such explicit solution,
but the expressions obtained for perturbative corrections
are much simpler than those found
from the equivalent Feynman diagrams.  Using these techniques, we were
able to extend the stochastic perturbation theory to the fourth order in the
noise strength.

Fourth order should be sufficient for most realistic calculations, but
does not provide enough information to determine the convergence properties
of the expansion, or determine eigenvalues beyond the first few.  
In this paper we develop
a third approach, based on construction of an explicit matrix
representation of the stochastic evolution operator. 
Numerical implementation requires a
truncation to finite dimensional matrices,
and is less elegant than the smooth conjugation method,
but for high expansion orders (here eighth, but higher orders seem quite
feasible) and many eigenvalues it is currently unsurpassed.
As with the previous formulations, it retains the periodic orbit structure,
thus inheriting valuable information about the dynamics.

In the following sections we define the stochastic dynamics and show how
to obtain matrix representations, both globally and located on the
periodic orbits, as an expansion in terms of the noise strength $\sigma$.
The matrix elements are obtained from derivatives of the dynamics computed
around each periodic orbit.  We give as a numerical example the quartic
map considered in both previous papers, although the approach is very
general and is by no means restricted to one dimension, to maps, or to
Gaussian noise.  We find that up to eighth order, the cumulants converge
super-exponentially  with the length of periodic orbit and the expansion is
now shown to be accurate to larger values of $\sigma$. 

\section{The stochastic evolution operator and its spectrum}
An individual trajectory in presence of additive noise is generated
by iteration
\beq
x_{n+1}=f(x_{n})+\sigma\xi_{n} 
\,,
\ee{mapf(x)-Diag}
where $f(x)$ is a map, 
$\xi_n$ a random variable with the normalized
distribution $p(\xi)$, 
and $\sigma$ parametrizes the noise strength.
In what follows we shall assume that the mapping $f(x)$ is
one-dimensional and expanding, and that the $\xi_n$ are uncorrelated.
A density of trajectories $\phi(x)$ evolves with time as
\beq
\phi_{n+1}(y) =
\left(
\Lop
\circ
\phi_{n}\right)(y)
= \int dx \, \Lop(y,x) \phi_{n}(x)
\ee{DensEvol}
where $\Lop$ is the {\evOper}
\bea
\Lop(y,x) &=& \delta_\sigma(y-f(x))
	\continue
  \delta_\sigma(x)
  &=& \int \delta(x-\sigma \xi) p(\xi) d\xi 
  \,=\, \frac{1}{\sigma} p\left( \frac{x}{\sigma} \right)
\,.
\label{oper-Diag}
\eea
For a repeller the leading eigenvalue of the {\evOper}
yields a physically measurable property of the dynamical system,
the escape rate from the repeller.
In the case of deterministic flows, the periodic orbit theory
yields explicit formulas for the spectrum of $\Lop$ as zeros
of its {\fd}\rf{QCcourse}. Our goal here
is to explore the extent to which such methods are applicable to
systems with noise and to quantum systems. In particular, we are
interested in exploring the dependence of the eigenvalues 
$\eigenvL(\sigma)$ of $\Lop$ 
on the noise strength parameter $\sigma$.

The eigenvalues are determined by the eigenvalue condition
\beq
\eigCond(\sigma,\eigenvL(\sigma)) = \det(1-\Lop/\eigenvL(\sigma)) =0
\label{eigCond}
\eeq 
where
$
\eigCond(\sigma,1/z) = \det(1-z\Lop)
$
is the {\fd} of  the {\evOper} $\Lop$.
Computation of such determinants
commences with evaluation of the traces of powers of
the {\evOper}
\bea
\tr {z\Lop \over 1-z\Lop}        
        &=&
         \sum_{n=1}^{\infty} C_{n} z^n
        \,,\qquad C_{n} = \tr \Lop^n
\,,
\label{tr-L-ith-Diag}
\eea
which are then used to compute the cumulants
$Q_{n}=Q_{n}(\Lop)$ in  the cumulant expansion
\beq
\det(1-z\Lop)			 
= 1- \sum_{n=1}^{\infty} Q_{n} z^n
\,,
\ee{Fred-cyc-exp-Diag}
by means of the recursion formula
\beq
Q_n = {1 \over n}\left( C_n -C_{n-1} Q_1 - \cdots C_1 Q_{n-1}\right)
\ee{Fd-cyc-exp-Diag}
which follows from the relation
\beq
        \det(1-z\Lop)
        \defeq
        \exp\left(-\sum_n^\infty {z^n \over n} \tr \Lop^n \right)
\,.
\ee{det-tr-Diag}
Our task is to compute the cumulants $Q_n$. 
We start by introducing a matrix representation for $\Lop$.

\section{Matrix representation of evolution operator}
\label{ENTIRE}

As the mapping $f(x)$ is expanding by assumption,  the {\evOper} \refeq{DensEvol}  smoothes the initial
distribution $\phi(x)$. Hence it is natural to assume that
the distribution $\phi_{n}(x)$ is analytic, and represent it as
a Taylor series, intuition being that the action of $\Lop$ will 
smooth out fine detail in initial distributions and the expansion
of $\phi_{n}(x)$ will be dominated by the leading terms in the series.

An analytic function $g(x)$ has a Taylor series expansion
\[
g(x) = \sum_{m=0}^\infty
  \frac{x^{m}}{m!} \left.  \frac{\pde^{m}}{\pde y^{m}}g(y) \right|_{y=0}
\,.  
\]
Expanding $\Lop(y,x)$ in Taylor series in $y$ 
enables us to rewrite traces of $\Lop^n$ as 
\bea
\tr \Lop^2
	&=&
 \int dx dy\, 
\Lop(y,x)
\Lop(x,y)
	\continue
	&=&
 \sum_{m,m'} \int dx dy\, 
  \left(\frac{y^{m'}}{m'!} 
  \left.  \frac{\pde^{m'}}{\pde v^{m'}}
	\Lop(v,x)
  \right|_{v=0}\right)
	\ceq  \qquad\qquad
  \left(\frac{x^{m}}{m!} 
  \left.  \frac{\pde^{m}}{\pde u^{m}}
	\Lop(u,y)
  \right|_{u=0}\right)
\nnu
\eea
Following H.H.~Rugh\rf{Rugh92} we
now define the matrix ($m,m'= 0,1,2, ...$)
\beq
 \left(\Lmat{}\right)_{m'm} = 
 \left. \frac{\pde^{m'}}{\pde y^{m'}}
  \int dx \, \Lop(y,x)
\frac{x^{m}}{m!} \right|_{y=0} .  
\ee{Lmat-Diag}
$ \Lmat{}$ is a matrix representation of $\Lop$ which 
maps the $x^m$ component of the density of trajectories $\phi_n(x)$ in \refeq{DensEvol} to the  $y^{m'}$ component of the density $\phi_{n+1}(y)$, with $y=f(x)$.
The desired traces can now be evaluated as traces of the matrix 
representation $ \Lmat{}$,
$
\tr \Lop^n
	=
\tr \Lmat{}^n
\,.
$
As $\Lmat{}$ is infinite dimensional, in actual
computations we have to truncate it to a given finite order.
The Feynman diagrammatic and the  smooth conjugation methods developed
in the preceding papers\rf{noisy_Fred,conjug_Fred}
require no such approximations. However, 
as we shall see below, for expanding flows the structure of $ \Lmat{}$
is such that its finite truncations give very accurate spectra.

Our next task is to evaluate the matrix elements of $ \Lmat{}$.

\section{Weak noise expansion of the evolution operator} 

We have written the operator $\Lop$ in \refeq{oper-Diag} in terms of
the Dirac delta function,
$
\Lop(x',x)=\int \delta(x'-f(x)-\sigma\xi)p(\xi)d\xi
$,
in order to emphasize that
in the weak noise limit the stochastic trajectories are concentrated 
along the deterministic trajectory $x' = f(x)$.
Hence it is natural to
expand the delta function in a Taylor series
in $\sigma$
\bea
\Lop(x',x) &=&
	\delta(x'-f(x)) 
	\ceq
	+\,
\sum_{n=2}^{\infty}\frac{(-\sigma)^n}{n!}\delta^{(n)}(x'-f(x))
\int \xi^n p(\xi)d\xi  
\,,
\nnu
\eea
where
$
	\delta^{(n)}(y) = {\pde^n \over \pde y^n} \delta(y)
	\,.
$
This yields a representation of the
{\evOper} centered along the deterministic trajectory, with 
the 
{\FPoper} $\delta(x'-f(x))$, and corrections given by derivatives
of delta functions weighted by  moments of the noise
distribution $a_n=\int p(\xi)\xi^nd\xi$, 
\begin{equation}
\Lop(x',x)
	=
	\delta(x'-f(x)) \,+\,
\sum_{n=2}^{\infty}\frac{(-\sigma)^n}{n!} a_n \delta^{(n)}(x'-f(x)).  
\label{opexp}
\end{equation}
In our  numerical tests we find it convenient to assume that the noise
is Gaussian,
$
p(\xi)=
e^{-\xi^2/2}/{\sqrt{2\pi}}
\,.
$
For the Gaussian noise all $a_n$ moments are known,
and the weak noise expansion of $\Lop$ is 
\bea
	\Lop(x',x)
	&=& 
	{1 \over \sqrt{2 \pi \sigma^2}} e^{-{(x'-f(x))^2/2\sigma^2} }
	\continue
	&=& 
	\sum_{n=0}^{\infty}
	\frac{\sigma^{2n}}{n!2^n} \delta^{(2n)}(x'-f(x))
	\continue
	&=& 
	\delta(x'-f(x)) + {\sigma^2 \over 2} \delta^{(2)}(x'-f(x)) 
		\ceq \qquad\qquad
	 + {\sigma^4 \over 8} \delta^{(4)}(x'-f(x)) + \cdots
	\,.
\label{delGaussExp}
\eea
The choice of Gaussian noise is not essential,
as the methods that we develop here apply
equally well to any other peaked smooth noise distribution, as
well as space dependent noise distributions $p(x,\xi)$. 
In any case, as the neighborhood of any trajectory is nonlinearly distorted by the flow, the  integrated noise is never Gaussian, but colored.

\section{Local matrix representation of evolution operator}
\label{LocMatr-Diag}

Traces of powers of the {\evOper} $\Lop^n$ are now also a power
series in $\sigma$, with contributions composed of 
$\delta^{(m)}(f(x_a)-x_{a+1})$ segments.
The contribution is non-vanishing only if the sequence
$x_1,x_2,...,x_n, x_{n+1} = x_{1}$ is a periodic orbit of
the deterministic map $f(x)$.
Thus  the series expansion of $\tr \Lop^n$ has support on
all periodic points $x_a = x_{a+n}$ of period $n$,
$f^n(x_a)=x_a$; the skeleton of periodic points of the deterministic
problem also serves to describe the weakly stochastic flows.
The contribution of the
$x_a$ neighborhood 
is best presented by introducing a coordinate system $\field_a$
centered on the  cycle points,
together with a notation for the
map \refeq{mapf(x)-Diag} and the operator \refeq{oper-Diag}
centered on the
$a$-th cycle point
\bea
x_a &\to& x_a+\field_a \,, \qquad a=1,...,n_p
	\continue
f_a(\field) &=& f(x_a+\field)
	\continue
{\Lop}_a(\field_{a+1},\field_a)
	    &=&
   \Lop(x_{a+1}+\field_{a+1},x_a+\field_a)
\,.
\eea
The weak noise expansion \refeq{opexp} for the $a$-th segment operator
is given by
\[
{\Lop}_a(\field',\field)=\sum_{n=0}^{\infty}\frac{(- \sigma)^{n}}{n!}
a_n \delta^{(n)}(\field'+x_{a+1}- f_a(\field))
\,.
\]

Repeating the steps that led to \refeq{Lmat-Diag}
we construct the local matrix representation of $\Lop_{a}$ centered on
the $x_a \to x_{a+1}$ segment of the deterministic trajectory
\bea
 \left(\Lmat{a}\right)_{m'm} 
	&=& 
 \left. \frac{\pde^{m'}}{\pde \field'^{m'}}
  \int d\field \,\Lop_{a}(\field',\field)
\frac{\field^{m}}{m!} \right|_{\field'=0} .  
	\continue
	&=& 
 \sum_{n=max(m-m',0)}^{\infty}
\frac{(-\sigma)^n}{n!}a_n(\Bmat{a})_{m'+n,m}
\,.
\label{BtoL}
\eea
Due to its simple dependence on the Dirac delta function,
$\Bmat{}$ can expressed in terms of derivatives of the inverse
 of $f_a(\field)$:
\bea
(\Bmat{a})_{n m}
	&=& 
	\left. \frac{\pde^{n}}{\pde \field'^{n}}
    \int d\field \, 
     \delta(\field' + x_{a+1} -f_a(\field))
    \frac{\field^{m}}{m!}
    \right|_{\field'=0}
	\continue
	&=& 
   \left. \frac{\pde^{n}}{\pde \field'^{n}} \frac{(f_a^{-1}(x_{a+1}+\field')-x_a)^{m}}{m!|f_a'(f_a^{-1}(x_{a+1}+\field'))|}\right|_{\field'=0}
        \continue
        &=&
   \frac{\mbox{sign}(f_a')}{(m+1)!}\left.\frac{\pde^{n+1}(
{\cal F}_a(\field')^{m+1})}{\pde \field'^{n+1}}\right|_{\field'=0},  
\label{Bmatrix}
\eea
 where we introduced the shorthand notation ${\cal F}_a(\field')=
f_a^{-1}(x_{a+1}+\field')-x_a$.

If we expand ${\cal F}_a(\field')$ in a Taylor series, the constant term
is
zero, since $f_a^{-1}(x_{a+1})=x_a$. So we can write:
\bea
{\cal
F}_a(\field')=\sum_{l=1}^{\infty}\frac{{\cal
F}_a^{(l)}}{l!}\field'^l,
\eea
where $1/{\cal F}_a^{(1)}=f'_a$.

The matrix elements can be calculated explicitly as a multinomial expansion \cite{abramo}
\bea
\left(\sum_{l=1}^{\infty}\frac{x_l}{l!}t^l\right)^m & = & m!\sum_{n=l}^{\infty}
\frac{t^n}{n!} \continue
\ceq \cdot \, \sum(n|a_1,...,a_n)'x_1^{a_1}...x_n^{a_n}, 
\label{Abram}
\eea
where the second sum $\left(\sum\right)$ goes over all non-negative integers
such that:
\bea
a_1+2a_2+...+na_n=n, \mbox{\hspace{0.3cm}} a_1+a_2+...+a_n=m,
\eea
and the multinomial coefficient is:
\bea
(n|a_1,a_2,...,a_n)'=\frac{n!}{(1!)^{a_1}a_1!(2!)^{a_2}a_2!...(n!)^{a_n}a_n!}.
\eea
We apply the formula $\refeq{Abram}$ to ${\cal F}_a(\field')$ with power $m+1$:
\bea
({\cal F}_a(\field'))^{m+1} &=& (m+1)! \sum_{l=m+1}^{\infty}\frac{\field'^n}{n!}\sum(l|a_1,a_2,...,a_l)' 
\continue
\ceq \mbox{} \cdot \, ({{\cal F}_a}^{(1)})^{a_1}({{\cal F}_a}^{(2)})^{a_2}...({{\cal F}_a}^{(l)})^{a_l}.
\eea
For the $(n+1)$ -th derivative of this expression evaluated at $\field'=0$ only the $l=n+1$ term is non-vanishing. 
The matrix elements vanish for $n<m$, so $\Bmat{}$ is
a lower triangular matrix:
\bea (\Bmat{a})_{n m} &=& \sum(n+1|a_1,a_2,...,a_{n+1})' 
\continue
\ceq \cdot \, ({{\cal F}_a}^{(1)})^{a_1}({\cal F}_a^{(2)})^{a_2}...({\cal F}_a^{(n+1)})^{a_{n+1}}.
\eea

The diagonal and the nearest off-diagonal matrix elements 
can  easily be worked out. Here we show the first four expressed in terms
 of the derivatives of the original map:

\bea
(\Bmat{a})_{m m}~~~ &=& \frac{1}{|f_a'|f_a'^{m}}
	\continue
(\Bmat{a})_{m+1,m} &=& - \frac{1}{2} \frac{(m+2)!}{m!}  \frac{f_a''}{|f_a'|f_a'^{m+2}}
	\label{Bexplicit}\\
(\Bmat{a})_{m+2,m} &=& 
	-\frac{(m+3)!}{24 m! |f_a'|f_a'^{m}}
\left(   \frac{f_a'''}{f_a'^{3}}       -3 (m+4) \frac{(f_a'')^2}{f_a'^{4}}  
\right)
	\continue
(\Bmat{a})_{m+3,m} &=& 
-	\frac{(m+4)!}{48 m!}{|f_a'|f_a'^{m}}
\left(  2 \frac{f_a''''}{f_a'^{4}} - 4(m+5) \frac{f_a'' f_a'''}{f_a'^{5}}
	\right.
	\ceq \qquad\qquad
	\left.
	+ (m+5)(m+6) \frac{f_a''^{3}}{f_a'^{6}} 
\right)
	\continue
& &\cdots
\nnu
\,,
\eea
where $f_a'$, $f_a''$,
$\cdots$ refer to the derivatives of $f(x)$ evaluated at the
periodic point $x_a$.

By assumption the map is expanding, $|f_a'|>1$. Hence
the diagonal terms drop off exponentially, as $1/|f_a'|^{m+1}$,
the terms below the diagonal fall off even faster, and 
we are justified in truncating $\Bmat{a}$, as truncating the
matrix to a finite one introduces only exponentially small errors.

In the local matrix approximation the traces of \evOper s are approximated by
\beq
\left. \tr {\Lop}^n\right|_{\mbox{\tiny saddles}}
=\sum_{p} n_p \sum_{r=1}^{\infty} \delta_{n, n_p r} \tr \Lmat{p}^r
= \sum_{j=0}^{\infty}C_{n j}\sigma^j
\,,
\label{tracenp}
\eeq
where
$\tr \Lmat{p}  = \tr{ \Lmat{\cl{p}}\Lmat{2}\cdots \Lmat{1}}$
is the contribution of the $p$ cycle, and the power series in
$\sigma^j$ follows from the expansion \refeq{BtoL} of $\Lmat{a}$
in terms of $\Bmat{a}$.  The subscript 
{\tiny saddles} is a reminder that this is a saddle-point
approximation to $\tr {\Lop}^n$ (see \refref{noisy_Fred} for a discussion),
valid as an asymptotic series in the limit of weak noise.

As a simple check of the above formulas, consider the noiseless case,
for which the $(\Lmat{a})_{m'm} = (\Bmat{a})_{m'm}$ matrices are 
a representation of the deterministic \FPoper\ $\left. \Lop \right|_{\sigma=0}$.
The $\Lmat{a}$  are triangular with diagonal elements
$ 
(\Lmat{a})_{m m}=  \frac{1}{|f_a'|f_a'^{m}}
\,.
$
The trace of the $\Lop$ on a periodic orbit $p$ is therefore
\[
\tr \Lmat{p}  
 	= \tr{ \Lmat{\cl{p}}\Lmat{2}\cdots \Lmat{1}}
	    =\sum_{m=0}^{\infty} \frac{1}{|\ExpaEig_p|\ExpaEig_p^{m}} 
	=\frac{1}{|1-\ExpaEig_p|}
\,,
\]
and we recover the standard
deterministic trace formula\rf{QCcourse} for the {\FPoper}
\beq
\tr{\Lop}^n=
\sum_p \cl{p} \sum_{r=1}^\infty
\delta_{n,\cl{p} r}
\frac{1}{|1-\ExpaEig_p^r|} 
\,.
\eeq

\section{Perturbative corrections to eigenvalues}
\label{trtoeig}

The eigenvalue condition \refeq{eigCond}
is an implicit equation for the eigenvalue $\eigenvL=\eigenvL(\sigma)$ 
of form $\eigCond(\sigma,\eigenvL(\sigma)) = 0$.
As the eigenvalue condition is satisfied for any $\sigma$,
all total derivatives of the eigenvalue condition with respect to $\sigma$
vanish, leading to
\bea
0 &=& 
	 {d \over d\sigma} \eigCond(\sigma,\eigenvL(\sigma ))
 = 
	{d \eigenvL \over d\sigma} {\pde \eigCond \over \pde \eigenvL}
		\,+\, 
	{\pde \eigCond \over \pde \sigma}
	\continue
0 &=& 
{d^2 \eigenvL \over d\sigma^2} 
 {\pde \eigCond \over \pde \eigenvL}
	     \,+\,
\left({d \eigenvL \over d\sigma}\right)^2
                    {\pde^2 \eigCond \over \pde \eigenvL^2}
	     \,+\, 2 {d \eigenvL \over d\sigma} 
			{\pde^2 \eigCond \over \pde \sigma  \pde \eigenvL}
	 \,+\,{\pde^2 \eigCond \over \pde \sigma^2}
\label{sec-derPert} \\
0 &=& 
{d^3 \eigenvL \over d\sigma^3} 
 {\pde \eigCond \over \pde \eigenvL}
	     \,+\,
3{d^2 \eigenvL \over d\sigma^2}
 {d \eigenvL \over d\sigma}
                    {\pde^2 \eigCond \over \pde \eigenvL^2}
	     \,+\,
\left({d \eigenvL \over d\sigma}\right)^3
                    {\pde^3 \eigCond \over \pde \eigenvL^3}
\ceq
	     \,+\, 3 {d^2 \eigenvL \over d\sigma^2} 
			{\pde^2 \eigCond \over \pde \sigma  \pde \eigenvL}
	     \,+\, 3 \left({d \eigenvL \over d\sigma}\right)^2 
	{\pde^3 \eigCond \over \pde \sigma  \pde \eigenvL^2}
\continue
\ceq     \,+\, 3 {d \eigenvL \over d\sigma}
 {\pde^3 \eigCond \over \pde \sigma^2  \pde \eigenvL}
	 \,+\,{\pde^3 \eigCond \over \pde \sigma^3}
\,,
\nnu
\eea
and so on. 
$\eigenvL(0)$ can be computed by cycle expansions for a deterministic,
noiseless flow. $\sigma \neq 0$ then parametrizes 
a weak perturbation to the deterministic 
\FPoper\ $\left. \Lop \right|_{\sigma=0}$. 
The above formulas enable us to compute recursively, order by order in
$\sigma^n$, the perturbative corrections to the eigenvalues of $\Lop$ 
\beq
\eigenvL(\sigma)
 = \sum_{m=0}^\infty \eigenvL_m  \sigma^m
\,,\qquad 
\eigenvL_m = {1 \over m!} \left.{d^m~ \over d\sigma^m}\eigenvL(\sigma) \right|_{\sigma=0}
\,,
\ee{pertExp1}
in terms of partial derivatives of the eigenvalue condition 
$\eigCond(\sigma,\eigenvL(\sigma))$
\beq
F_{kl} = \left.{\pde^{k+l}~~~ \over  \pde \eigenvL^k \pde \sigma^l}
			\eigCond(\sigma,\eigenvL)
	 \right|_{\sigma=0, \eigenvL = \eigenvL(0)}
\,.
\eeq
In this notation the formulas  \refeq{sec-derPert} for $\eigenvL_m$
take the form
\bea
\eigenvL_1 &=& - {F_{01} \over F_{10}}
	\continue
\eigenvL_2 &=& -{1 \over 2 \, F_{10}}\left( F_{02} + 2 \,  F_{11} \, \eigenvL_1 + 2 \, F_{20} \, \eigenvL_1^{2}  \right)
	\label{EigPert-Diag}\\
\eigenvL_3 &=& -{1 \over 3! \, F_{10}} 
	\left(
		 F_{01} + 3 \, F_{12} \, \eigenvL_1 + 6 \, F_{11} \, \eigenvL_2 
	\right.
	\ceq \qquad\qquad
	\left.
		+ \, 3 \, F_{21} \, \eigenvL_1^{2} + 6 \, F_{20} \, \eigenvL_1 \, \eigenvL_2 + F_{30} \, \eigenvL_1^{3}  \right)
\,.
\nnu
\eea
As shown in \refref{QCcourse}, $F_{kl}$ 
can be computed from explicit cycle expansions.
However, in numerical calculations
we find it more expedient to proceede by first expressing
the {\fd} $F$ in terms of the cumulants.
The traces of $\Lmat{}^n$ evaluated by \refeq{BtoL} yield a series in
$\sigma^j$,  and
the $\sigma^j$ coefficients $Q_{n j}$ in the cumulant expansion 
\beq
F=\det(1-z\Lop) = 1-\sum_{n=1}^{\infty}\sum_{j=0}^{\infty}Q_{n j}z^n\sigma^j
\label{qum}
\eeq
are then obtained recursively  from the traces, as in \refeq{Fd-cyc-exp-Diag}:
\beq
Q_{n j} = \frac{1}{n}\left(C_{n j}
           - \sum_{k=1}^{n-1}\sum_{l=0}^{j}Q_{k,j-l}C_{n-k,l}
                     \right)
\,.
\ee{Q-Sig-expan}
This gives $F = F(z = 1/\eigenvL \, , \sigma)$ and the partial derivatives $F_{kl}$ can be found.
Substituted in \refeq{EigPert-Diag} they yield the perturbative corrections to the eigenvalues. The above calculations can be efficiently done by manipulating formal Taylor series.

\section{Numerical tests}

Here we continue the calculations of the eigenvalue corrections
described in  \refrefs{noisy_Fred,conjug_Fred},
where more details and discussion may be found.  
We test our perturbative expansion on the repeller
of the 1-dimensional map
\beq 
f(x)=20\left(\frac{1}{16}-\left(\frac{1}{2}-x\right)^4\right)
\,.
\ee{testQuartic}
This repeller is a clean example of an ``Axiom~$A$'' expanding
system of bounded nonlinearity and complete binary symbolic dynamics, 
for which the deterministic \evOper\ eigenvalues
converge super-exponentially with the cycle length\rf{Rugh92}.

We start the numerical calculations by determining 
all prime cycles up to a given length. 
For each prime cycle $p$  we compute
the truncated evolution matrix   $\Lmat{p}$ and its repetitions  $\Lmat{p}^r$
to the given order in $\sigma$, and evaluate the traces \refeq{tracenp}.
For the map at hand we find that truncations
of size [$16\times 16$] suffice to achive double precision accuracy 
for most cycles. However, as the short orbits are less unstable, they require
larger matrix truncations in order to attain the same precision, and 
we employ a [$28\times 28$] truncation for the $2$-cycles, and a
[$34\times 34$] truncation for the fixed points. 
With the coefficients in the traces expansion \refeq{tracenp}
evaluated numerically, the cumulants and the perturbative eigenvalue
corrections follow from \refeq{Q-Sig-expan} and \refeq{EigPert-Diag}.
In case at hand, a good first approximation is obtained already at $n=2$
level, using only 3 prime cycles,
and $n=6$ (23 prime cycles in all) is in this example sufficient
to exhaust the limits of double precision arithmetic. 

\begin{figure}[hbt]
\centerline{\psfig{figure=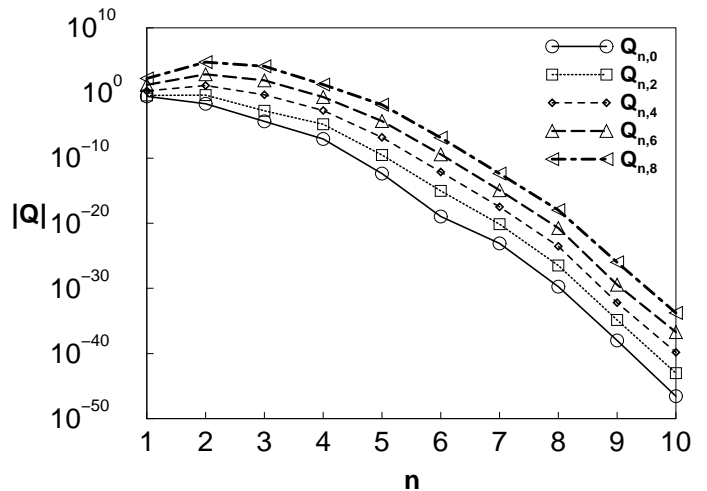,width=9cm}}
\caption{The perturbative corrections 
\refeq{Q-Sig-expan} to the cumulants $Q_{n j}$ plotted as a function
of cycle length $n$ (for perturbation orders $j=0,2,4,6,8$) all exhibit
super-exponential convergence.}
\label{qumfig}
\end{figure}

The size of the cumulants is indicated in \reffig{qumfig},
and the perturbative corrections to the leading eigenvalue 
of the weak-noise \evOper\ are given in \reftab{tabPert}. 
Encouragingly, the  value of 
$\eigenvL_6 = 2076.47\ldots$
computed here is not  wildly different to  our previous numerical 
estimate\rf{conjug_Fred} of 
2700. 
Both the cumulants and the eigenvalue corrections exhibit a
super-exponential convergence with the truncation cycle length $n$.
The super-exponential convergence has been proven for the deterministic,
$\eigenvL_0$ part of the eigenvalue\rf{Rugh92}, but the proof
has not been extended to the stochastic evolution operators.

We have chosen to test the formalism on 
this simple example, as here we are in a fortunate situation that
the  escape rate for arbitrary noise strength $\sigma$ can  be
calculated numerically by other methods to a rather high accuracy. 
For example, one can discretize the stochastic kernel on a 
spatial lattice\rf{noisy_Fred} and determine numerically 
the leading eigenvalue.

\begin{figure}[hbt]
\centerline{\psfig{figure=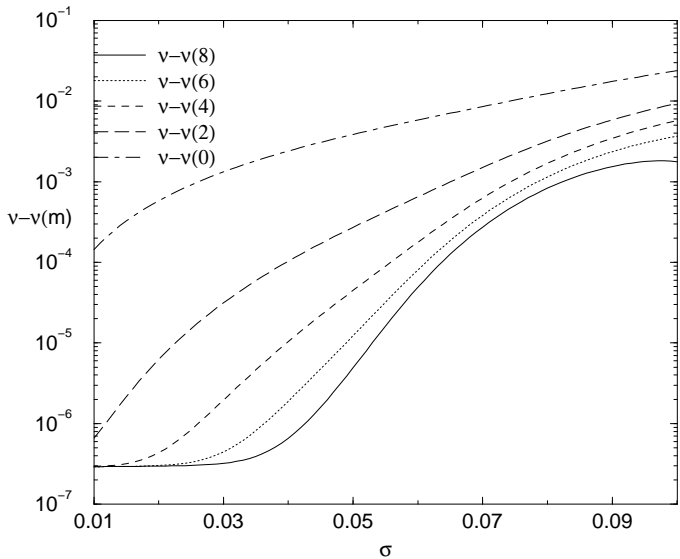,width=9cm}}
\caption{
The difference between the numerical and perturbative eigenvalue
$|\eigenvL(\sigma) - \eigenvL(m,\sigma)|$.
The plateau at $10^{-7}$ is a numerical artifact 
due to the limited accuracy of the lattice discretization calculation. 
\label{nufig2}}
\end{figure}

The perturbative result in terms of periodic orbits and the weak noise corrections
is  compared to the eigenvalue computed by the numerical
lattice discretizationin \reffig{nufig2}, with the absolute difference between
the numerical and the $m$th order
perturbative results plotted.
We see that the perturbative result
$\eigenvL(m,\sigma)=\sum_{k=0}^{m/2} \eigenvL_{2k}\sigma^{2k}$
indeed improves as more perturbative terms are added.

\section{Summary and outlook}

In this paper we study  evolution of a classical dynamical system
with additive noise. In the limit of weak noise the traces of 
the corresponding {\evOper} are approximated by sums of local
traces computed on periodic orbits. 
Here we present a new, computationally efficient technique for
evaluation of these local traces based on  
a matrix representation of the {\evOper}, and show that method is
powerful enough to enable us to 
compute 2 more orders of perturbation theory.

The local matrix representation can be interpreted as follows.
Substituting \refeq{tracenp} into \refeq{det-tr-Diag} we obtain
\bea
\left. 
 \det(1 - z \Lop)
\right|_{\mbox{\tiny saddles}} 
	=
 \prod_{p} \det(1 - z^{n_p} \Lmat{p}) 
\,.
\label{fredholmprod}
\eea
In other words, in 
the saddle-point approximation the spectrum of the {\em global}
{\evOper} $\Lop$
is in this approach
pieced together from the {\em local} spectra computed cycle-by-cycle
on neighborhoods of individual prime cycles with periodic boundary
conditions. Vattay\rf{VatBS} was first to formulate the $\hbar$
corrections to the semi-classical Gutzwiller theory in terms of local
spectra.
Here we have shown that also the stochastic flows can be suspended
on the skeleton of classical periodic orbits in this way. 

With so many orders of perturbation theory, we are now
poised to address the issues raised by the asymptotic series nature
of perturbative expansions.  We can now hope to resum the series to all
orders, making use of techniques such as the Borel resummation, the asymptotic
expansions of general integrals of saddlepoint type, and 
asymptotics beyond all orders\rf{Dingle}. 
All of this is beyond the scope of
the present paper, and we defer a full discussion of asymptotics 
to a forthcoming paper\rf{asym_Fred}.

\section{Acknowledgements}

G.V. and G.P. gratefully acknowledges the financial support of
the Hungarian Ministry of Education, FKFP 0159/1997, OMFB, OTKA T25866/F17166.
G.V. thanks
Bruno Eckhardt  the  cordial hospitalty at the Department of Physics of
the Philipps-Universit\"at Marburg and the Humboldt Fundation for
support. 
G.P. thanks the EU network ``Pattern formation, noise and 
spatio-temporal chaos in complex systems'', TMR contract
ERBFMRXCT960085, for partial 
support. N.S. is supported by the Danish Research Academy Ph.D.
fellowship.


\onecolumn
\begin{table*}
\begin{tabular}{clllll}
$n$&$\eigenvL_0$&$\eigenvL_2$&$\eigenvL_4$&$\eigenvL_6$&$\eigenvL_8$\\\hline
1&0.308&0.42&~2.2                                    &17.4             &168.0    \\
2&0.37140&1.422&32.97                                &1573.3           &112699.9 \\
3&0.3711096&1.43555&36.326                           &2072.9           &189029.0 \\
4&0.371110995255&1.435811262&36.3583777              &2076.479         &189298.8 \\
5&0.371110995234863&1.43581124819737&36.35837123374  &2076.4770492     &189298.12802\\
6&0.371110995234863&1.43581124819749&36.358371233836 &2076.47704933320 &189298.128042526\\
\hline
\end{tabular}
\caption{Significant digits of the leading deterministic eigenvalue
$\eigenvL_0$, and the
$\sigma^2, \cdots,  \sigma^{8}$ perturbative coefficients
\refeq{pertExp1}, calculated from the
cumulant exapansion of the {\fd},
as a function of the cycle truncation length $n$.
Note the super-exponential convergence of all coefficients. 
\label{tabPert}}
\end{table*}

\end{document}